\documentstyle[preprint,aps]{revtex}
\renewcommand{\theequation}{\thesection.\arabic{equation}}
\begin{document}
\draft
\preprint{Preprint No. IP--BBSR--96/33}
\title{An Explicit and Simple Relationship Between Two Model Spaces. }
\author{J. S. Prakash{\thanks {jsp@iopb.ernet.in}}} 
\address{Institute Of Physics\\ Sachivalaya Marg, Bhubaneswar 751 005, India}
\date{May 1996}
\maketitle

\begin{abstract} 
An explicit and simple correspondence, between the basis of the
model space of $SU(3)$ on one hand and that of $SU(2)\otimes
SU(2)$ or $SO(1,3)$ on the other, is exhibited for the first
time.  This is done by considering the generating functions for
the basis vectors of these model spaces.
\end{abstract}
PACS number(s): 

\newpage

\section{\bf Introduction}

In this paper our concern is with special types of infinite
dimensional Hilbert spaces known as model spaces for group
representations.  Therefore they are spaces on which one builds
the irreducible representations(irreps) of groups.  They are
spaces of functions, with a well defined inner product, such
that they contain every irrep of the given group exactly once
when these functions are restricted to a suitable homogeneous
space under the group action.  Such spaces have been with us for
a long time now the best example being that of the group
$SU(2)$.  The model space of this group is the infinite
dimensional space spanned by the monomials in two complex
variables as the basis vectors.  We can decompose this space
into a direct sum of unitary irreps of $SU(2)$ by using the
Bargmann\cite{BV} inner product.  Similary one knows how to
construct the representations of $SO(3)$ in the space of all
square integrable functions on the $2$-sphere.  This gives a
model of the representations of $SO(3)$.  Because of this nice
property model spaces attracted the attention of mathemacians
and physicists who carried out detailed investigations on how to
construct them.  Gelfand and his coworkers\cite{Gelf1} initiated
a systematic investigation of constructing models for every
connected reductive group.  Biedenharn and Flath\cite{LCBDF}
constructed a model of the Lie algebra $sl(3,C)$ and also found
that the action of $sl(3, C)$ on this model extends to an action
of the larger Lie algebra $so(8, C)$.  Following this Gelfand
and Zelevinskij constructed models of representations of
classical groups\cite{Gelf2,Kramer,Gelf3}.  Now relating the
basis states of any two given Hilbert spaces to each other is
agreeably a difficult proposition.  But since model spaces are
apparently very nice spaces one expects the task of finding
suitable maps under which the basis states of one space are
related to the basis states of the other to be a less formidable
one.  The purpose of this paper is to confirm this optimism by
explicitly exhibiting the isomorphism between the model space
for the finite dimensional unitary representations of the group
$SU(3)$ on one hand and the model space for the finite
dimensional unitary representations of the group $SU(2)\otimes
SU(2)$ or for the finite dimensional nonunitary representations
of the group $SO(1, 3)$ on the other.  We achieve this by making
use of the generating function for the model space basis of the
group $SU(3)$ written by us
\cite{jsphss,JSPdiff,JSPWEYL,JSPWIGD} for
the first time and the generating function for the model space
basis of the group $SU(2) \otimes SU(2)$ written by
Schwinger\cite{SJ} long time ago.  Incidentally it is
interesting to note that one of these groups, namely $SU(3)$, is
an internal symmetry group whereas the other group, $SO(1,3)$, is
a space-time symmetry group.

The plan of the paper is as follows.  In sections 2 and 3 we
briefly review some basic results which lead us to the
generating functions of the groups $SU(3)$, $SO(1,3)$ and
$SU(2)\otimes SU(2)$.  We then describe the relationship between
the model spaces of $SU(3)$ and of $SU(2)\otimes SU(2)$ or
$SO(1,3)$ in section 4.  The last section is devoted to a
discussion of our results.

\section{Review of some relevant results on $SU(3)$ }

In this section we briefly review some results concerning the
groups $SU(3)$, $SU(2)\otimes SU(2)$ and $SO(1,3)$ which lead us
to the generating functions of the basis states of the model
spaces of these groups.  The details of these results can be
found in \cite{jsphss}.  

$SU(3)$ is the group of $3\times 3$ unitary unimodular matrices
$A$ with complex coefficients. It is a group of $8$ real
parameters.  The matrix elements satisfy the following
conditions

\begin{eqnarray}
A&=&(a_{ij})\, , \quad A^\dagger A = AA^\dagger
= I,\, \quad \mbox{where}\,\, I\,\, \mbox{is the identity matrix
and,\,}\, \quad \mbox{det}(A)=1\, .
\label{A}
\end{eqnarray}

\subsection{Parametrization using $\vec Z$ and $\vec W$}.

One well known parametrization of $SU(3)$ is due to Murnaghan
{\cite{mfd}}, see also {\cite{bmarh}}.  But here we use a
parametrization of $SU(3)$ as a complex unit spherical cone.  That
is we now give a parametrization\cite{bmarh} of $A\in SU(3)$
in terms of the complex variables $z_1, z_2, z_3$ and
$w_1,w_2,w_3$ corresponding to the irreps ${\underline 3}$ and
${\underline 3^*}$.

For this purpose we constrain these variables to the
intersection of the two unit $5$-spheres

\begin{eqnarray}
\vert{z_1}\vert^2+\vert{z_2}\vert^2+\vert{z_3}\vert^2=1\, ,\qquad
\vert{w_1}\vert^2+\vert{w_2}\vert^2+\vert{w_3}\vert^2=1\, ,
\end{eqnarray}
with the complex cone
\begin{eqnarray}
z_1w^{*}_1+z_2w^{*}_2+z_3w^{*}_3=0\, .
\label{paracone}
\end{eqnarray}

\noindent Then $A\in SU(3)$ can be written as below

\begin{eqnarray}
A=\pmatrix{z_1^* & z_2^* & z_3^* \cr w_1^* & w_2^* & w_3^* \cr
u_1 & u_2 & u_3} \, ,
\end{eqnarray}
where
\begin{eqnarray}
u_i=\sum_{j,k}\epsilon_{ijk}z_jw_k\, .
\end{eqnarray}

The following two points are to be noted. (i) This unit cone is
a homogeneous space\cite{bmarh} for the action of the group
$SU(3)$ and (ii) the group manifold itself can be idetified with
this cone.  This is contrary to the popular belief that only in
the case of $SU(2)$ the group manifold can be identified with a
geometric surface.  More importantly this cone serves as a model
space for the irreps of $SU(3)$.

\subsection{A realization of the Lie algebra of $SU(3)$}

The following is a realization of the Lie algebra of the group
$SU(3)$ in terms of the variables $z_i,\, w_i$

\begin{eqnarray}
\pi^0 &=&(z_1\frac{\partial }{\partial z_1} - z_2\frac{\partial }{\partial z_2}
- w_1\frac{\partial }{\partial w_1} + w_2\frac{\partial
}{\partial w_2})\, ,\,\,
\pi^-=(z_2\frac{\partial }{\partial z_1}-w_1\frac{\partial
}{\partial w_2})\, ,\,\,
\pi^+=(z_1\frac{\partial }{\partial z_2}-w_2\frac{\partial
}{\partial w_1})\, ,\nonumber \\
K^-&=&(z_3\frac{\partial }{\partial z_1})\, ,\,\,
K^+=(z_1\frac{\partial }{\partial z_3}+{\bar z_3}(z_1w_1+z_2w_2)\frac{\partial }{\partial w_1}
+{\bar z_3}^2z_1\frac{\partial }{\partial w_1})\, , \nonumber \\
K^0&=&(z_3\frac{\partial }{\partial z_2})\, ,\,\,
{\bar {K^0}}=(z_2\frac{\partial }{\partial z_3}+{\bar
z_3}(z_1w_1+z_2w_2) \frac{\partial }{\partial w_2}-{\bar z_3}^2
\frac{\partial }{\partial {\bar z_3}})\, ,\nonumber \\
\eta &=&(z_1\frac{\partial }{\partial z_1}+z_2\frac{\partial }{\partial z_2}
-2z_3\frac{\partial }{\partial z_3}-w_2\frac{\partial }{\partial w_2}
-w_1\frac{\partial }{\partial w_1}+2{\bar z_3}\frac{\partial
}{\partial {\bar z_3}})\, .
\end{eqnarray}

It is clear from the above that in this realization the
generaters in the pairs $\pi^+$ and $\pi^-$, $K^+$ and $K^-$,
and $K^0$ and ${\bar {K}}^0$ are not adjoints of each other.
Using an 'auxiliary' measure and by requring that their
representative matrices to be adjoints of each other we can
compute the 'true' normalizations of our basis states.\cite{jsphss}

\subsection{Irreducible Representations - The Constraint.}

Tensors constructed out of these two $3$ dimensional
representations span an infinite dimensional complex vector
space.

If we now impose the constraint

\begin{eqnarray}
 z_1w_1+z_2w_2+z_3w_3=0\, ,
\label{z.w}
\end{eqnarray}
on this space we obtain an infinite dimensional complex vector
space in which each irreducible representation of $SU(3)$ occurs
once and only once.  Such a space is called a {\em model space}
for $SU(3)$.  Further if we solve the constraint
$z_1w_1+z_2w_2+z_3w_3=0$ and eliminate one of the variables, say
$w_3$, in terms of the other five variables $z_1, z_2, z_3, w_1,
w_2$ we can write a genarating function to generate all the
basis states of all the irreps of $SU(3)$.  This generating
function is computationally a very convenient realization of the
basis of the model space of $SU(3)$.  Moreover we can define a
scalar product on this space by choosing one of the variables,
say $z_3$, to be a planar rotor ${\exp}(i\theta)$.  Thus the
model space for $SU(3)$ is now a Hilbert space with this scalar
product between the basis states.  Our basis states are
orthogonal with respect to this scalar product but are not
normalized.  The 'true' normalizations can be computed using
this scalar product by requiring that the irreps of $SU(3)$ be
unitary.  The above construction was carried out in detail in a
previous paper by us {\cite{jsphss}}.  For easy accessability we
give a self-contained summary of those results here.

Before going to the next section we note that the equation for
the constraint Eq.(\ref{z.w}), used to construct the irreducible
representations, is slightly different from the one used in the
parametrization Eq.(\ref{paracone}).  But, since we are not
going to work with the group invariant measure, resulting from
our parametrization of $SU(3)$, we need not worry about this
fact.

\subsection{Explicit realization of the basis states}
 
\noindent {\bf{(i) {Generating function for the basis states of $SU(3)$}}}

The generating function for the basis states of the irreps of
$SU(3)$ can be written as
 
\begin{equation}
g(p,q,r,s,u,v)={\exp}(r(pz_1+qz_2)+s(pw_2-qw_1)+uz_3+vw_3)\, .
\label{g()}
\end{equation}

The coefficient of the monomial $p^Pq^Qr^Rs^Su^Uv^V$ in the
Taylor expansion of Eq.(\ref{g()}), after eliminating $w_3$
using Eq.(\ref{z.w}), in terms of these monomials gives the
basis state of $SU(3)$ labelled by the quantum numbers $P, Q, R,
S, U, V$.

\noindent {\bf{{ (ii) Formal generating function for the basis
states of $SU(3)$ }}}

The generating function Eq.(\ref{g()}) can be written formally
as 

\begin{equation}
g=\sum_{P,Q,R,S,U,V} p^Pq^Qr^Rs^Su^Uv^V \vert PQRSUV)\, ,
\label{FGF}
\end{equation}
where $\vert PQRSTUV)$ is an unnormalized basis state of $SU(3)$
labelled by the quantum numbers $P,Q,R,S,U,V$.
----
Note that the constraint $P+Q=R+S$ is automatically satisfied in
the formal as well as explicit Taylor expansion of the generating
function.

\subsection{Labels for the basis states}

\noindent {{\bf{(i) Gelfand-Zetlein labels}}}\\

Normalized basis vectors are denoted by,
$\vert{M,N;P,Q,R,S,U,V}>$.  All labels are non-negative
integers.  All Irreducible Represenatations (IRs) are uniquely
labeled by $(M, N)$.  For a given IR $(M, N)$, labels
$(P,Q,R,S,U,V)$ take all non-negative integral values subject to
the constraints:

\begin{equation}
R+U=M\hspace{.1in},\hspace{.1in} S+V=N\hspace{.1in},\hspace{.1in} P+Q=R+S.
\end{equation}

The allowed values can be presribed easily: $R$ takes all values
from $0$ to $M$, and $S$ from $0$ to $N$.  For a given $R$ and
$S$, $Q$ takes all values from $0$ to $R+S$.\\

\noindent {{\bf{(ii) Quark model labels}}}\\

The relation between the above Gelfand-Zetlein labels and the
Quark Model labels is as given below.

\begin{eqnarray}
2I&=&P+Q=R+S, \,\, 2I_3=P-Q,\nonumber\\
Y&=& \frac{1}{3} (M-N) + V-U \nonumber\\
&&= \frac{2}{3} (N-M)-(S-R)\, .\nonumber\\
\end{eqnarray}
where as before $R$ takes all values from $0$ to $M$.  $S$ takes all
values from $0$ to $N$.  For a given $R$ and $S$, $Q$ takes all
values from $0$ to $R+S$.

\subsection{'Auxiliary' scalar product for the basis states}

\noindent{\bf{Notation}}

Hereafter, for simplicity of notation we assume, all variables
other than the $z^i_j$ and $w^i_j$ where $i, j=1,,2,3$ are real
eventhough we have treated them as comlex variables at some
places.  Our results are valid even without this restriction as
we are interested only in the coefficients of the monomials in
these real variables rather than in the monomials themselves.

The scalar product to be defined in this section is 'auxiliary'
in the sense that it does not give us the 'true' normalizations
of the basis states of $SU(3)$.  However it is computationally
very convenient for us as all computations with this scalar
product get reduced to simple Gaussian integrations and the
'true' normalizations themselves can then be got quite easily.  

{\bf{(i) Scalar product between generating functions of basis
states of $SU(3)$}}

We define the scalar product between any two basis states in
terms of the scalar product between the corresponding generating
functions as follows :

\begin{eqnarray}
(g', g)&=& {\int_{-\pi}^{+\pi}}{\frac{d\theta}{2\pi}} \int
\frac{d^{2}z_1}{\pi^2} \frac{d^{2}z_2}{\pi^2}
\frac{d^{2}w_1}{\pi^2} \frac{d^{2}w_2}{\pi^2}
{\exp}(-\bar{z_1}z_1 - \bar{z_2}z_2 - \bar{w_1}w_1
-\bar{w_2}w_2)\nonumber\\
&&\nonumber \\
&&\times {\exp}((r'(p'z_1+q'z_2) + s'(p'w_2-q'w_1) - \frac{-v'}{z_3}
(z_1w_1 + z_2w_2) + u'\bar{z}_3) \nonumber \\
&&\nonumber\\
&&\times {\exp}((r(pz_1 + qz_2) + s(pw_2-qw_1) - \frac{-v}{z_3}
(z_1w_1 + z_2w_2) + uz_3)\, , \nonumber \\
&&\nonumber \\
&=& (1-v'v)^{-2} \left (\sum_{n=0}^{\infty}
\frac{(u'u)^n}{(n!)^2}\right )
{\exp}\left [(1-v'v)^{-1}(p'p + q'q)(r'r + s's)\right ]\, .
\label{gg'}
\end{eqnarray}

{\bf{(ii) Choice of the variable $z_3$}}

To obtain the Eq.(\ref{gg'}) we have made the choice
\begin{eqnarray}
z_3=\exp(i\theta )\, .
\label{z3}
\end{eqnarray}

The choice, Eq.(\ref{z3}), makes our basis states for $SU(3)$
depened on the variables $z_1,z_2,w_1,w_2$ and $\theta $.

\subsection{Normalizations}

\noindent {\bf{{(i) 'Auxiliary' normalizations of unnormalized
basis states}}} 

The scalar product between two unnormalized basis states,
computed using our 'auxiliary scalar product, is given by,

\begin{eqnarray}
M(PQRSUV)&\equiv &(PQRSUV\vert PQRSUV)
=\frac{(V+P+Q+1)! }{P! Q! R! S! U! V! (P+Q+1)}\, .
\label{M}
\end{eqnarray}

\noindent {\bf{(ii) Scalar product between the unnormalized 
and normalized basis states}}

The scalar product, computed using our 'auxiliary' scalar
product, between an unnormalized basis state and a normalized
one is given by the next equation where it is denoted by
$(PQRSUV\vert\vert PQRSUV>$.

\begin{equation}
(PQRSUV\vert\vert PQRSUV>=N^{-1/2}(PQRSUV)\times M(PQRSUV)\, .
\label{(||)}
\end{equation}

\noindent {\bf{{(iii) 'True' normalizations of the basis
states}}} 

We call the ratio of the 'auxiliary' norm of the unnormalized
basis state represented by $\vert PQRSUV)$, and the scalar product
of the unnormalized basis state with a normalized
Gelfand-Zeitlin state, represented by $\vert PQRSUV > $, as
'true' normalization.  It is given by

\begin{eqnarray}
N^{1/2}(PQRSUV)&\equiv & \frac{(PQRSUV\vert
PQRSUV)}{<PQRSUV\vert PQRSUV>}\nonumber \\ 
&&\nonumber\\
&&=\left ( \frac{(U+P+Q+1)! (V+P+Q+1)! }{P! Q! R!S! U! V!
(P+Q+1)}\right )^{1/2}\, .
\label{N}
\end{eqnarray}

\section{Review of results on the groups $So(1,3)$ and $SU(2)\otimes SU(2)$}

We make use of the results contained in Schwinger's
work\cite{SJ} for this purpose.  Moreover we do so in the
Bargmann representation of the boson creation and annihilation
operators.  Therefore introduce the operators

\begin{eqnarray}
z_\zeta&=&(z_1,\, z_2),\quad {\frac{\partial
}{\partial z_\zeta}} = (\frac{\partial }{\partial z_1},\,
\frac{\partial }{\partial z_2}),\quad
w_\zeta=(w_1,\, w_2),\qquad {\frac{\partial
}{\partial w_\zeta}} = (\frac{\partial }{\partial w_1},\,
\frac{\partial }{\partial w_2})\, ,
\end{eqnarray}
obeying the following commutation relations

\begin{eqnarray}
[\frac{\partial}{\partial z_\zeta},\, \frac{\partial}{\partial
z_{\zeta^{'}}}]=0,\qquad [z_\zeta,\,
z_{\zeta^{'}} ]=0,\qquad [\frac{\partial}{\partial
z_\zeta},\, z_{\zeta^{'}}] = \delta_{\zeta\zeta^{'}}\, ,\nonumber\\
\end{eqnarray}

\begin{eqnarray}
[\frac{\partial}{\partial w_\zeta},\, \frac{\partial}{\partial
w_{\zeta^{'}}}]=0,\, \qquad [w_\zeta,\,
w_{\zeta^{'}} ]=0,\, \qquad [\frac{\partial}{\partial
w_\zeta},\, w_{\zeta^{'}}]=\delta_{\zeta\zeta^{'}}\, ,
\end{eqnarray}
where $z$ and $w$ are two complex variables.

Then the following operators

\begin{eqnarray}
{\cal J}_{1+}&=&(z_1\frac{\partial }{\partial z_2})\, ,\quad 
{\cal J}_{1-}=(z_2\frac{\partial }{\partial z_1})\, ,\quad
{\cal J}_{13}={1 \over 2}(z_1\frac{\partial }{\partial z_1} 
- z_2\frac{\partial }{\partial z_2}),\,
\label{j1}
\end{eqnarray}
\begin{eqnarray}
{\cal J}_{2+}&=&(w_1\frac{\partial }{\partial w_2}),\,\quad
{\cal J}_{2-}=(w_2\frac{\partial }{\partial w_1}),\,\quad 
{\cal J}_{23}={1 \over 2}(w_1\frac{\partial }{\partial w_1} 
- w_2\frac{\partial }{\partial w_2})\, ,
\label{j2}
\end{eqnarray}
obey the commuation relations of the ordinary angular momentum algebra.

And the operators given below

\begin{eqnarray}
{\cal I_+}&=&(z_1\frac{\partial }{\partial w_1}+z_2\frac{\partial
}{\partial w_2})\, ,\quad 
{\cal I_-}=(w_1\frac{\partial }{\partial z_1}+w_2\frac{\partial
}{\partial z_2})\, ,\nonumber\\
{\cal I}_3&=&{1 \over 2}(z_1\frac{\partial }{\partial z_1} + z_2\frac{\partial
}{\partial z_2} - w_1\frac{\partial }{\partial w_1} + w_2\frac{\partial
}{\partial w_2})\, ,\quad
{\cal K_+}=(z_1w_2 - z_2w_1)\, ,\nonumber\\
{\cal K_-}&=&(\frac{\partial }{\partial z_1}\frac{\partial
}{\partial w_2} + \frac{\partial }{\partial z_2}\frac{\partial
}{\partial w_1})\, ,\quad 
{\cal K}_3={1\over 2}[(z_1\frac{\partial }{\partial
z_1} + z_2\frac{\partial}{\partial z_2} + w_1\frac{\partial
}{\partial w_1} + w_2\frac{\partial}{\partial w_2})+1]\, ,
\end{eqnarray}
form the Lie algebra of the group $SO(1,3)$ as one can verify that

\begin{eqnarray}
[ {\cal {I}}_{3}, \, {\cal {I}}_{\pm} ] &=& {\pm} {\cal
{I}}_{\pm},\, \qquad
[{\cal {I}}_{+}, \, {\cal I}_{-} ] = 2 {\cal I}_{3}\,, 
\label{Icomm}
\end{eqnarray}
\begin{eqnarray}
[{\cal K}_{3},\, {\cal K}_{\pm} ]  &=& {\pm} {\cal K}_{\pm},\,\qquad
[{\cal K}_{+},\, {\cal K}_{-} ] =-2 {\cal K}_{3} \, ,
\end{eqnarray}
and that the two sets of operators commute with each other.

In a similar fashion one can define the Lie algebra of the group
$SU(2)\otimes SU(2)$ in terms of the following operators,

\begin{eqnarray}
{\cal J_+}&=&(z_1\frac{\partial }{\partial z_2} + w_1\frac{\partial
}{\partial w_2})\, ,\quad 
{\cal J_-}=(z_2\frac{\partial }{\partial z_1}+w_2\frac{\partial
}{\partial w_1})\, ,\nonumber\\
{\cal J}_3&=&{1 \over 2}(z_1\frac{\partial }{\partial z_1} - z_2\frac{\partial
}{\partial z_2} + w_1\frac{\partial }{\partial w_1} - w_2\frac{\partial
}{\partial w_2})\, ,
\end{eqnarray}
together with operators ${\cal I_+}$,\, ${\cal I_-}$,\, and
${\cal I}_3$ obeying the following commutation relations

\begin{eqnarray}
[{\cal I}_3,\, {\cal I_\pm} ]=\pm{\cal I_\pm} ,\qquad [{\cal
I_+},\, {\cal I_-} ]=2{\cal I}_3\nonumber
\end{eqnarray}
\begin{eqnarray}
[{\cal J}_3,\, {\cal J_\pm }]=\pm{\cal J_\pm},\qquad [{\cal
J_+},\, {\cal J_-} ]=2{\cal K}_3\, .
\end{eqnarray}

As in the previous case these two sets of operators also commute
with each other.

\subsection{Generating functions for the basis states of the 
groups $SO(1,3)$ and $SU(2)\otimes SU(2)$}

\noindent {\bf{{ (i) Explicit generating function for the basis
states of $SU(2)\otimes SU(2)$ or of $SO(1,3)$}}}

Here we will be concerned only with the finite dimensional
representations of the group $SO(1,3)$.  As this group is 
non-compact these representations are non-unitary.  They can be got
by taking the direct products of the irreps of the finite
dimensional unitary irreps of the group $SU(2)$.  As with all
direct product groups these are the irreps of the group
$SU(2)\otimes SU(2)$ also.  Below we describe the generating
function for the basis states of these irreps.

Denote the generating function for the basis states of the
groups $SU(2)\otimes SU(2)$ or $SO(1,3)$ by $g_{SO(1,3)
}$.  Then this generating function is given by\cite{SJ}

\begin{eqnarray}
g_{SO(1,3)}=\exp (v(z_1w_2 - z_2w_1) + r(pz_1 + qz_2) + s(pw_1 + qw_2)\, .
\label{gso13}
\end{eqnarray}

If we take the $z_i,\, w_i$ as creation operators then the above
generating function acts on a vaccum state $\psi_0$.

The coefficient of the monomial $p^Pq^Qr^Rs^Sv^V$ in the Taylor
expansion of Eq.(\ref{gso13}) gives the basis state of
$SU(2)\otimes SU(2)$ or that of $SO(1,3)$ labelled by the
quantum numbers $P, Q, R, S, V$.

\noindent {\bf{{(ii) Formal generating function for the basis
states of $SU(2)\otimes SU(2)$ or of $SO(1,3)$}}}

The generating function Eq.(\ref{gso13}) can be written formally
as 

\begin{equation}
{g}=\sum_{P,Q,R,S,V} p^Pq^Qr^Rs^Sv^V \vert PQRSV>\, ,
\label{Fgso13F}
\end{equation}
where $\vert PQRSV>$ is a normalized basis state of
$SU(2)\otimes SU(2)$ or of $SO(1,3)$ labelled by the quantum
numbers $P,Q,R,S,V$.  This is in contrast to the case of
$SU(3)$ in which case the corresponding basis state is
unnormalized.

\subsection{Scalar product between the basis states}

Schwinger\cite{SJ} had calculated the scalar product between the
basis states which is given as follows

\begin{eqnarray}
({g}^{'}_{SO(1,3)},\,\, {g}_{SO(1,3)}) = &=&
(1-v'v)^{-2} {\exp}\left [(1-v'v)^{-1}(p'p +
q'q)(r'r + s's)\right ]\, .
\label{2ndgg'}
\end{eqnarray}

\subsection{Correspondence with the usual labels}

In terms of the usual angular momentum labels our labels
$P,Q,R,S,V$ can be expressed as follows 

\begin{eqnarray}
P &=& j   + m\, , \quad Q = j - m\, ,\quad R=j + j_1 - j_2\, ,\quad
S = j_2 + j   - j_1\, ,\quad V = j_1 + j_2 - j
\end{eqnarray}
and vice-versa. We also note that the constraint $P+Q=R+S$
holds.

Solving for the angular momentum quantum numbers we get,

\begin{eqnarray}
j &=& {{P+Q} \over 2},\,\quad m = {{P-Q} \over 2}\, ,\quad
j_1 = {{R+V} \over 2}\, ,\quad j_2 = {{S+V} \over 2}\, .
\end{eqnarray}

We conclude that the basis states given by this generating
function are labelled by the eigenvalues of $J_3$, ${\cal I}_3$
and ${\cal K}_3$ that is by $m=m_1+m_2$, $\mu =j_1 - j_2$ and
$\nu= j_1 + j_2 +1$.\cite{SJ}  It is clear that the basis states can be
equivalently labelled by the quantum numbers $j_1,\, j_2,\, j,\,
m$ or by $j_1,\, j_2,\, m_1,\, m_2$.  Here $J_3=J_{13} + J_{23}$.  

Our generating function can be obtained from the more usual
generating function which gives basis states labelled by the
quantum numbers $j_1,\, j_2,\, m_1,\, m_2$ by operating with the
differential operator\cite{SJ}

\begin{eqnarray}
\exp \left (v \left [ \frac{\partial }{\partial
t_1}\frac{\partial }{\partial t_2}\right ] + r\left ( x
\frac{\partial }{\partial t_1}\right ) + s \left ( x
\frac{\partial }{\partial t_2} \right ) \right )
\end{eqnarray}
on 
\begin{eqnarray}
\exp (t_1(z) + t_2(w))
\end{eqnarray}
where the $x$, $z$ and $w$ are the two component vectors $(p,\,
q)$, $(z_1,\, z_2)$ and $(w_1,\, w_2)$.  The derivatives are to
be evaluated at $t_1=t_2=0$ and the square bracket have the
following meaning
\begin{eqnarray}
[zw]=z_1w_2-z_2w_1
\end{eqnarray}

\section{The Correspondence Between the Model Spaces}

Now we are in a position to take a look at the relationship
between the model spaces of $SU(3)$ and $SU(2)\otimes SU(2)$ or
of $SO(1,3)$.

For this we write below the generating functions for the basis states
of these groups and compare them\cite{jsphss,SJ}.

\begin{eqnarray}
{g}_{SO(1,3)}&=&{\exp}(r(pz_1+qz_2)+s(pw_1+qw_2)+v(z_1w_2
- z_2w_1)\nonumber\\ &&=\sum^\infty_{2j=0}\sum_{j_1+j_2=j} ({2j! \over
(j_1+j_2-j)!(j+j_1-j_2)!(j_2+j-j_1)!
}r^{j+j_1-j_2}s^{j_2+j-j_1}v^{j_1+j_2-j}\nonumber\\ &&\times
(z_1w_2-z_2w_1)^{j_1+j_2-j}\cdot (pz_1+qz_2)^{j+j_1-j_2} \cdot
(pw_1+qw_2)^{j2j-j1}\, ,
\label{1ggFso13}
\end{eqnarray}

and
\begin{eqnarray}
{g}_{SU(3)} &=&{\exp}(r(pz_1+qz_2)+s(pw_2-qw_1)-{v\over
z_3}(z_1w_1+z_2w_2)+uz_3)\nonumber \\
&&=\sum^\infty_{2j=0}\sum_{j_1+j_2=j} ({2j! \over
(j_1+j_2-j)!(j+j_1-j_2)!(j_2+j-j_1)!
}r^{j+j_1-j_2}s^{j_2+j-j_1}v^{j_1+j_2-j}\nonumber\\ &&\times
(-z_1w_1-z_2w_2)^{j_1+j_2-j}\cdot (pz_1+qz_2)^{j+j_1-j_2} \cdot
(pw_2-qw_1)^{j2j-j1}\cdot z^{(U+j-j_1-j_2)}_3\, .
\label{1ggF}
\end{eqnarray}

From the expressions for the formal generating functions
Eqs.(\ref{FGF},\ref{Fgso13F}) for the groups at hand we recall that
coefficients of the monomials $p^Pq^Qr^Rs^Su^Uv^V$ in the
expansion of the generating function for $SU(3)$ and
coefficients of the monomials $p^Pq^Qr^Rs^Sv^V$ in the expansion
for the generating function for $SO(1,3)$ or $SU(2)\otimes
SU(2)$ are the basis functions for the various finite
dimensional irreps of these groups.  This means that given any
set of five positive integers $P,Q,R,S,V$ a monomial
$p^Pq^Qr^Rs^Sv^V$ can be associated with it in the expansion of
each of these two generating functions.  But in the case of
$SU(3)$ there is an additional factor of $u^U$ multiplying these
monomials with $U$ taking any arbitrary positive integral power.
Therefore there is a many-to-one correspondence between the
terms of the power series expansion of these generating
functions.  This correspondence in turn leads to a many-to-one
correspondence between the coefficients of these monomials which
are our basis functions.  That is, all those basis states of the
group $SU(3)$ with the same quantum numbers $P,Q,R,S,V$ but with
different quantum numbers $U$ are in a many-to-one
correspondence with the basis states of the group $SO(1,3)$(or
of $SU(2)\otimes SU(2)$) which have the same quantum numbers
$P,Q,R,S,V$.  This establishes the relationship between the
basis states of these model spaces.  Below we will work out the
precise map by which we can get the basis states of the latter
group(s) from those of the former and note a few interesting
points about this relationship.

Now let us look at the expressions for the scalar products
between the basis states of these two groups Eqs.(\ref{gg'},\ref{2ndgg'}).  We note
that if we ignore the $\exp (uz_3)$ part of the generating
function for the basis states of $SU(3)$ then the scalar product
between the basis states of $SU(3)$ is identical with the scalar
product between the basis states of $SO(1,3)$(or of
$SU(2)\otimes SU(2)$.  But from the expressions for the
'auxiliary' and 'true' normalizations for the basis
states Eqs.(\ref{M},\ref{N}) of $SU(3)$ we know that with respect to this
'auxiliary' scalar product these basis states are orthogonal but
are not normalized.  The 'true' normalization having been
computed\cite{jsphss} using the unitarity of the group
representation matrices in these bases.  On the otherhand with
respect to this scalar product the basis states of the IRs of
$SO(1,3)$(or of$SU(2)\otimes SU(2)$ are not only orthogonal but
are also normalized.  Thus using their generating functions we
can relate the othogonal but not normalized basis states of
$SU(3)$ with the othonormal basis of the other group(s).  

Next let us work out the precise relationship between the
generating functions.  

The map

\begin{eqnarray}
z_1\rightarrow z_1,\, \qquad z_2\rightarrow z_2, \,
\qquad w_1\rightarrow w_2 \qquad w_2\rightarrow -w_1,\, 
\end{eqnarray}
followed by a multiplication by $z^{(U+j-j_1-j_2)}_3$, with $U$
being any positive integer, takes the coefficients of the
monomials, in the expansion of the generating function for the
group $SO(1,3)$ or $SU(2)\times SU(2)$, that is the basis
functions of these groups, onto the coefficients of these
monomials in the expansion of the generating function for the
group $SU(3)$ , that is onto its basis functions.  It should be
noted that if instead of the groups $SO(1,3)$ or $SU(2)\otimes
SU(2)$ we consider the groups $SO(1,3)\otimes $(Planar Rotor
Group) or $SU(2)\otimes SU(2)\otimes$ (Planar Rotor Group) then the
above prescription of multiplication by $z^{(U+j-j_1-j_2)}_3$,
with $U$ being any positive integer, can be dropped.

Ignoring the $\exp (uz_3)$ part of the generating function for the
basis states of $SU(3)$ the inverse of the above map is the
following

\begin{eqnarray}
z_1\rightarrow z_1,\, \qquad z_2\rightarrow z_2,\, 
\qquad w_2\rightarrow w_1 \qquad -w_1\rightarrow w_2\qquad
z_3\rightarrow 1\, .
\end{eqnarray}

Since under the above described maps the generating functions
for the basis states of the groups at hand are related we
conclude that the individual basis states also get related to
each other under the same maps.

There is a word of caution about this mapping.  It should be
clearly borne in mind that this is a mapping between the basis
states of one model space and the basis states of another.  It
is {\em not} a mapping between the irreducible multiplets of one
group into those of the other.  In other words, though a single
basis state of one space is mapped to a single basis state of
the other space, in general a single irreducble multiplet in one
is {\em not} mapped into a single multiplet of the other.  

Since these groups, $SU(3)$ on one hand and $SO(1,3)$(or
$SU(2)\otimes SU(2)$, are not in a group subgroup relationship
to each other the correspondence that we worked out between
their basis states is not covered by either group subduction or
by group induction.  

\noindent{\bf Examples}

The following are some examples illustrating the effect of the
the mapping mentioned above which takes the basis states of
$SU(3)$ into those of $SO(1,3)$(or of $SU(2)\otimes SU(2)$.
Here we have labelled the irreps of $SU(3)$, standing to the
left of the equations below, by their dimensions and those of
$SU(2)\otimes SU(2)$(or of $SO(1,3)$ by the values of $j_1,\,
j_2$.  For details see the appendix.

\begin{eqnarray}
\underline 3&=&({\underline{1\over 2}} \otimes {\underline 0})
\oplus ({\underline 0} \otimes {\underline 0})\, ,\nonumber\\
\nonumber\\
\underline{3}^{*}&=&({\underline{0}}\otimes{\underline{1\over 2}})
\oplus ({\underline{1\over 2}} \otimes {\underline{1\over
2}})_{j=0,\, m=0}\, ,\nonumber\\
\nonumber\\
\underline{8}&=&({\underline{1\over 2}}\otimes{\underline{1\over 2}})
\oplus ({\underline{1}} \otimes {\underline{1\over
2}})_{j={1\over 2}}\oplus ({\underline 0}\otimes {\underline
{1\over 2}})\, .
\end{eqnarray}

It is wellknown that the Casimir operators and their eigenvalues
of groups other than $SU(2)$ do not have any physical
interpretation (similar to the angular momentum).  In this
context it may be useful to make use of the above described
mapping to obtain algebraic expressions for the labels of irreps
of $SU(3)$, which can be related to the eigenvalues of the
Casimir operators of $SU(3)$, in terms of known physical quantum
numbers.  As is wellknown the irreps of the group $SU(3)$ are
labelled by two positive integers denoted by us by $M, N$.  Now
from the previous discussion we know that

\begin{eqnarray}
M=j+j_1-j_2-U\, ,\quad N=2j_2
\label{MN}
\end{eqnarray}
where $j_1, j_2$ and $j$ are the eigenvalues of the casimir
operators of the angular momentum algebras in
Eqs.(\ref{j1}, \ref{j2}) and in Eq.(\ref{Icomm}) and $U$ is the
eigenvalue of the planar rotor $e^{i\theta U}$.  Thus each irrep
of $SU(3)$ can be labelled by a quartet of angular momentum
labels $(j_1,\,j_2,\,j,\,U)$ instead of the usual two integers
$(M,\, N)$.

\section{Discussion}

In this paper we have established an explicit and simple
correspondence between the basis states of the irreps of $SU(3)$
and those of $SO(1,3)$ or of $SU(2)\otimes SU(2)$.  For this
purpose we have made use of the generating functions for the
basis states of the model spaces of these groups.  Thus in
general if one can write down generating functions for the basis
states of the model spaces of the basis states of various groups
then it may be easy to relate the basis states of different
groups to each other.  

We have made use of the relationship between the basis states of
$SU(3)$ and those of $SO(1,3)$ or of $SU(2)\otimes SU(2)$ to
relate the quantum numbers labelling these states.  This was
useful for us to obtain algebraic expressions for the labels of
the irreps of the group $SU(3)$ interms of angular momentum
quantum numbers.  

\newpage

\renewcommand{\theequation}{A.\arabic{equation}}
\setcounter{equation}{0}

\newpage

{\large\bf Appendix : Examples}\\

The following are some exmaples of the way that the members of some
of the multiplets of $SU(3)$ split, under the mapping discussed
in this paper, into basis states of $SO(1,3)$(or of
$SU(2)\otimes SU(2)$.  The $SU(3)$ states are correctly
normalized but the states resulting from the mapping are not
correctly normalized.  One can compute these normalizations also
using the scalar product Eq.(\ref{2ndgg'}).  But we have not shown it
here.

\begin{center}
$\underline{3}$($M$=1, $N$=0)\\
\vspace{0.5cm}
\begin{tabular}{|c|c|c|c|c|c|c|c|c|c|c|c|}\hline
. & $P$ & $Q$ & $R$ & $S$ & $U$ & $V$ & $I$   & $I_3$  & $Y$ & $\vert PQRSUV)$ & $N^{1/2}$ \\ \hline 
$u$ & 1 & 0 & 1 & 0 & 0 & 0 & 1/2 & $1/2$ & 1/3 & $z_1$ & $\sqrt{2}$ \\ \hline 
$d$ & 0 & 1 & 1 & 0 & 0 & 0 & 1/2 & $-1/2$ & 1/3   & $z_2$   & $\sqrt{2}$ \\ \hline 
$s$ & 0 & 0 & 0 & 0 & 1 & 0 & 0   &  0   & -2/3  & $z_3$     & $\sqrt{2}$ \\ \hline 
\end{tabular}
\end{center}
\vspace{0.3cm}

\hspace{7cm}$\Downarrow$

\vspace{0.3cm}

\begin{center}
$\underline 3$=$({\underline{1\over 2}} \otimes {\underline 0})
\oplus ({\underline 0} \otimes {\underline 0})$\\
$\underline{3}$($j+j_1-j_2+U$=$M$=1, $2j_2$=$N$=0)\\
\vspace{0.5cm}
\begin{tabular}{|c|c|c|c|c|c|c|c|c|c|c|c|}\hline
.   &$P$&$Q$&$R$&$S$& $V$&$j$  &$m$   &$j_1$&$j_2$&$\vert PQRSV)$ & $N^{1/2}$ \\ \hline 
 &1  & 0 & 1 & 0 &  0 & 1/2 &$1/2$ &1/2  &$0$  &$z_1$           & $\sqrt{2}$ \\ \hline 
 &0  & 1 & 1 & 0 &  0 & 1/2 &$-1/2$&1/2  &$0$  &$z_2$     & $\sqrt{2}$ \\ \hline 
\end{tabular}
\end{center}

\newpage

\begin{center}
$\underline{3}^*$($M$=0, $N$=1)\\
\vspace{0.5cm}
\begin{tabular}{|c|c|c|c|c|c|c|c|c|c|c|c|}\hline
.         & $P$ & $Q$ & $R$  & $S$  & $U$ & $V$ & $I$   & $I_3$  & $Y$    & $\vert PQRSUV)$ & $N^{1/2}$ \\ \hline 
$\bar{d}$ & 1   & 0   &  0   &  1   &  0 & 0 &   1/2    & $1/2$  & -1/3   & $w_2$            & $\sqrt{2}$ \\ \hline 
$\bar{u}$ & 0   & 1   &  0   &  0   &  0 & 0 &   1/2    & $-1/2$ & -1/3    & -$w_1$            & $\sqrt{2}$ \\ \hline 
$\bar{s}$ & 0   & 0   &  0   &  0   &  0 & 1 &   0      &  0     & 2/3      & $w_3$            & $\sqrt{2}$ \\ \hline 
\end{tabular}
\end{center}

\vspace{0.3cm}

\hspace{7cm}$\Downarrow$

\vspace{0.3cm}

\begin{center}
$\underline{3}^{*}=({\underline{0}}\otimes{\underline{1\over 2}})
\oplus ({\underline{1\over 2}} \otimes {\underline{1\over
2}})_{j=0,\, m=0}$\\
${\underline{3}^*}$($j+j_1-j_2+U$=$M$=0, $2j_2$=$N$=1)\\
\vspace{0.5cm}
\begin{tabular}{|c|c|c|c|c|c|c|c|c|c|c|c|}\hline
.   &$P$&$Q$&$R$&$S$&$V$&$j$&$m$   &$j_1$&$j_2$&$\vert PQRSV)$ & $N^{1/2}$ \\ \hline 
 &1  & 0 &0  &1  & 0 &1/2&$1/2$ &0    &1/2 &$w_1$            & $\sqrt{2}$ \\ \hline 
 &0  & 1 &0  &0  & 0 &1/2&$-1/2$&0    &1/2 &$w_2$           & $\sqrt{2}$ \\ \hline 
 &0  & 0 &0  &0  & 1 &0  &0     &1/2  &1/2 &$z_1w_2-z_2w_1$            & $\sqrt{2}$ \\ \hline 
\end{tabular}
\end{center}

\newpage

\begin{center}
$\underline{8}$($j+j_1-j_2+U=M$=1, $2j_2=N$=1)\\
\vspace{0.3cm}
\begin{tabular}{|c|c|c|c|c|c|c|c|c|c|c|c|}\hline
.            & $P$ & $Q$ & $R$ & $S$ & $U$ & $V$ & $I$  & $I_3$  & $Y$ & $\vert PQRSUV)$ & $N^{1/2}$   \\ \hline 
$\pi^+$      & 2   & 0   &  1  &  1  &  0  & 0   &  1   &  1     & 0   & $z_1w_2$         & $\sqrt{6}$  \\ \hline 
$\pi^0$      & 1   & 1   &  1  &  1  &  0  & 0   &  1   &  0     & 0   & $-z_1w_1+z_2w_2$ & $\sqrt{12}$ \\ \hline 
$\pi^-$      & 0   & 2   &  1  &  1  &  0  & 0   &  1   & -1     & 0   & $-z_2w_1$        & $\sqrt{6}$  \\ \hline 
$K^+$        & 1   & 0   &  1  &  0  &  0  & 1   &  1/2 &  1/2   & 1   & $z_1w_3$         & $\sqrt{6}$  \\ \hline
$K^0$        & 0   & 1   &  1  &  0  &  0  & 1   &  1/2 &  -1/2  & 1   & $z_2w_3$         & $\sqrt{6}$  \\ \hline
$\bar{K}^0 $ & 1   & 0   &  0  &  1  &  1  & 0   &  1/2 &  1/2   & -1  & $w_2z_3$	  & $\sqrt{6}$  \\ \hline
$K^-$        & 0   & 1   &  0  &  1  &  1  & 0   &  1/2 &  -1/2  & -1  & $-w_1z_3$ 	  & $\sqrt{6}$  \\ \hline
$\eta $      & 0   & 0   &  0  &  0  &  1  & 1   &  0   &  0     & 0   & $(z_3w_3=-z_1w_1-z_2w_2)$ & 2   \\ \hline
\end{tabular}
\end{center}

\vspace{0.3cm}

\hspace{7cm}$\Downarrow$

\vspace{0.3cm}

\begin{center}
$\underline{8}=({\underline{1\over 2}}\otimes{\underline{1\over 2}})
\oplus ({\underline{1}} \otimes {\underline{1\over
2}})_{j={1\over 2}}\oplus ({\underline 0}\otimes {\underline {1\over 2}})$\\
\vspace{0.3cm}
\begin{tabular}{|c|c|c|c|c|c|c|c|c|c|c|c|}\hline
.   &$P$&$Q$&$R$&$S$&$V$&$j$& $m$    &$j_1$&$j_2$& $\vert PQRSV)$ & $N^{1/2}$   \\ \hline 
 &2  &0  & 1 & 1 &0  &  1  &  1   &1/2  &1/2   & $z_1w_2$         & $\sqrt{6}$  \\ \hline 
 &1  &1  & 1 & 1 &0  &  1  &  0   &1/2  &1/2   & $z_1w_2+z_2w_1$ & $\sqrt{12}$ \\ \hline 
 &0  &2  & 1 & 1 &0  &  1  & -1   &1/2  &1/2   & $z_2w_2$        & $\sqrt{6}$  \\ \hline 
 &1  &0  & 1 & 0 &1  &  1/2&  1/2 &1    &1/2   & $z_1(z_1w_2-z_2w_1)$         & $\sqrt{6}$  \\ \hline
 &0  &1  & 1 & 0 &1  &  1/2&  -1/2&1    &1/2   & $z_2(z_1w_2-z_2w_1)$         & $\sqrt{6}$  \\ \hline
 &1  &0  & 0 & 1 &0  &  1/2&  1/2 &0    &1/2   & $w_1$	  & $\sqrt{6}$  \\ \hline
 &0  &1  & 0 & 1 &0  &  1/2&  -1/2&0    &1/2   & $w_2$ 	  & $\sqrt{6}$  \\ \hline
 &0  &0  & 0 & 0 &1  &  0  &  0   &1/2  &1/2   & $(z_1w_2-z_2w_1)$ & 2   \\ \hline
\end{tabular}
\end{center}

\end{document}